# Is Quantum Mechanics needed to explain consciousness ?

Knud Thomsen

Paul Scherrer Institute, 5232 Villigen-PSI, Switzerland

**Abstract**

In this short comment to a recent contribution by E. Manousakis [1] it is argued that the reported agreement between the measured time evolution of conscious states during binocular rivalry and predictions derived from quantum mechanical formalisms does not require any direct effect of QM. The recursive consumption analysis process in the Ouroboros Model can yield the same behavior.

One answer to the question posed in the title seems trivial and affirmative: as our total classical world is grounded in quantum mechanics' realm, so are brains and the processes they implement. More interesting is the issue whether quantum mechanical effects are directly responsible for the working of brains and in the generation of conscious perceptions.

In a recent contribution E. Manousakis claimed that experimental observations during binocular rivalry would provide support for a model according to which consciousness evolves as a quantum system including collapses of the wave function [1]. E. Manousakis proposes that in analogy to a quantum system in superposition "potential consciousness" evolves between "projections or measurements", which eliminate all but one outcome. Combining quantum formalism with time constants obtained in psychophysical experiments the author is able to fit selected experimental data [2].

There is no doubt that the reported agreement is remarkably good, but there is also no reason to infer from this that conditions of consciousness have to be equated with genuine quantum mechanical states. What can be asserted only is that some peculiar experimental results can be reproduced with a formalism akin to standard quantum mechanics. Any process featuring a suitable interplay between distinct phases with non-linear switching can successfully be described with the same formulas.

Rhythms with rather different frequencies, some of them most probably nested, abound in the human brain. Conscious content appears to be binned into a grained time structure, delimited by about 30 Hz at the short end and 3 seconds as the longest duration of a perceptual unit [3]. Overall, the dynamical behavior and the occurring time scales can with relative ease be understood in terms of the macroscopic and classical (as distinct from purely quantum mechanical) characteristics of diverse neurons and their connections [4].
Given the macroscopic time spans observed during binocular rivalry in the order of seconds there is no need for resorting to quantum mechanical effects to obtain fitting values.

One model which can yield the reported behavior and which can partly be described with E. Manousakis' formalism, is the Ouroboros Model [5]. It features a basic algorithmic structure for efficient minds. The following initially unconscious activity cycle is identified:

    ... anticipation,
    action / perception,
    evaluation,
    anticipation,...

Extensive evidence in the literature tells that all concepts are organized and stored as schemata, i.e. frames connecting specific constituents. In the Ouroboros Model activation at a time of part of a schema biases the whole structure and, in particular, missing features, thus triggering expectations. An iterative recursive monitor process termed 'consumption analysis' is checking how well such expectations fit with successive activations.

According to the Ouroboros Model, during phases of 'data collection' different, possibly mutual exclusive, frames are activated; the system has not settled on one attribution. When everything fits into a coherent and consistent scheme, consumption analysis records the success and the selected interpretation, whereas the gathering of new data continues. Due to the boundary conditions dictated during evolution for avoiding endless considerations, the process ends in any case after a finite time and the best then available interpretation is accepted.

In binocular rivalry, where it is not possible to arrive at one all-encompassing coherent perception, a second round would find less support from the features 'consumed' just before and instead salient activations that have not been integrated; consequently the perception will switch after a certain amount of new evidence has accumulated. The implicated thresholds are modulated by many influences, ranging from random noise and the habituation of detectors to biases exerted by the context and attention as well as health and intoxication factors [2,6,7,8,9].

Interrupting the visual stimuli to the eyes for some time disrupts the perception, and the process only continues as soon as its input is available again [10]. Thus, the effect of "freezing" of the temporal evolution can also be understood in the framework of the Ouroboros Model.

Note that the expendability of direct quantum mechanical effects in explaining the observed behavior during binocular rivalry is independent of the arguments whether the internal working of the brain can at all rely directly on genuine quantum mechanical processes [11,12].

The point here simply is that truly classical and macroscopic systems can embody algorithms which appear to mimic quantum mechanical effects and thus can be described to some extent using the same formalism. The Ouroboros Model offers one proposal how the reported perceptions and their alternations during binocular rivalry could come about.

Just the same as for the paper by E. Manousakis, nothing of the above makes it easy to understand that activity in the brain "collapsed" into one single pattern should entail conscious awareness [1]. A detailed account of how the Ouroboros Model comprising Higher Order Personality Activation, HOPA, gives rise to subjective consciousness will be presented elsewhere [13].

Returning to the original question asked in the title the answer advocated here is: **no**.